\documentclass[12pt]{article}
\usepackage{subfloat}
\pdfoutput = 1
\usepackage{graphics}
\usepackage{graphicx} %Include figure filesusepackage{graphicx} %Include figure files
\usepackage{multirow}
\usepackage{hhline}
\begin{document}
\date{Today}
%%%%%%%%%%%%%%%%%%%%
\title{{\bf{\Large Holographic free energy and thermodynamic geometry }}}
%%%%%%%%%%%%%%%%%%%%

\author{
{\bf {\normalsize Debabrata Ghorai}$^{a}$
\thanks{debanuphy123@gmail.com, debabrataghorai@bose.res.in}},\,
{\bf {\normalsize Sunandan Gangopadhyay}$^{b,c,d}
$\thanks{sunandan.gangopadhyay@gmail.com, sunandan@iiserkol.ac.in,  sunandan@associates.iucaa.in}}\\
$^{a}$ {\normalsize  S.N. Bose National Centre for Basic Sciences,}\\{\normalsize JD Block, 
Sector III, Salt Lake, Kolkata 700098, India}\\[0.2cm]
$^{b}$ {\normalsize Indian Institute of Science Education and Research, Kolkata}\\{\normalsize Mohanpur, Nadia 741246, India}\\ 
$^{c}$ {\normalsize Department of Physics, West Bengal State University, Barasat 700126, India}\\
$^{d}${\normalsize Visiting Associate in Inter University Centre for Astronomy \& Astrophysics,}\\
{\normalsize Pune 411007, India}\\[0.1cm]
}
\date{}

\maketitle

\begin{abstract}
{\noindent We analytically obtain the free energy and thermodynamic geometry of holographic superconductors in $2+1$-dimensions. The gravitational theory in the bulk dual to this $2+1$-dimensional strongly coupled theory lives in the $3+1$-dimensions and is that of a charged $AdS$ black hole together with a massive charged scalar field. The matching method is applied to obtain the nature of the fields near the horizon using which the holographic free energy is computed through the gauge/gravity duality. The critical temperature is obtained for a set of values of the matching point of the near horizon and the boundary behaviour of the fields. The thermodynamic geometry is then computed from the free energy of the boundary theory. From the divergence of the thermodynamic scalar curvature, the critical temperature is obtained once again. We then compare this result for the critical temperature with that  obtained from the matching method. }
\end{abstract}
\vskip 1cm

%%%%%%%%%%%%%%%%%%%%%%%%%%%%%%%%% Introduction %%%%%%%%%%%%%%%%%%%%%%%%%%%%%%%%%%%%%%%%%%%%%%%%%%%%
\section{Introduction}
There has been an immense amount of interest in studying strongly coupled systems using one of the most fascinating developments in modern theoretical physics, the gauge/gravity correspondence. The correspondence \cite{adscft1}-\cite{adscft4} gives an exact duality between a gravity theory living in a $(d+1)$-dimensional $AdS$ spacetime with a field theory having conformal invariance sitting on the $d$-dimensional boundary of this spacetime. In recent years, it was first demonstrated in \cite{hs2} that the formation of a scalar field condensate near the horizon of a black hole is possible below a certain critical temperature for a charged $AdS$ black hole coupled minimally to a complex scalar field. The gauge/gravity duality then implies that the scalar operator dual to the scalar field in the bulk acquires a non-zero vacuum expectation value in the boundary field theory. This is what is known as the holographic superconductor phase transition \cite{hs3}-\cite{hs6}. Thereafter, a lot of work has been done to study the properties of the holographic fermi liquid \cite{nw1},\cite{nw2}, holographic insulator/superconductor phase transition \cite{nw3}, transport properties of holographic superconductor both in the probe limit and away from the probe limit, that is including the effects of back reaction of matter fields on the background spacetime \cite{hs16}-\cite{dg1}. \\
Another interesting development that has taken place recently is the association of a geometrical structure with thermodynamic systems in equilibrium. This was first realized through the works in \cite{gs1}-\cite{gs3}. It was shown that one can get a Riemannian metric with an Euclidean signature from the equilibrium state of a thermodynamic system. The Riemannian scalar curvature can then be computed and captures the details of interactions of the thermodynamic system. It turns out that this framework based on a geometrical structure gives a handle to study critical phenomena \cite{gs3}.\\
In this paper, we set out to study the properties of $2+1$-dimensional holographic superconductors using the formalism of the thermodynamic geometry. We employ the matching method \cite{hs8}, \cite{fth1} to obtain the behaviour of the matter fields near the horizon of the black hole. This in turn is used to compute the critical temperature and the condensation operator. We obtain the critical temperature for a set of values of the matching point where the near horizon and boundary values of the fields are matched. The analysis is carried out for two sets of boundary conditions for the condensation operator, namely $\psi_{-}\neq 0$, $\psi_{+}=0$ and $\psi_{+}\neq 0$, $\psi_{-}=0$. The choice $\psi_{+}=0$ essentially implies that $\psi_{-}$ is dual to the expectation value of the condensation operator at the boundary. The requirement $\psi_{+}=0$ is necessary because we want the condensate to turn on without being sourced. We then proceed to compute the free energy of this $2+1$-dimensional holographic superconductor. The trick here is to relate the free energy of the theory on the boundary to the value of the on-shell action of the Abelian-Higgs sector of the full Euclidean action with proper boundary terms \cite{fth2},\cite{fth3}. From this, we compute the thermodynamic metric using the formalism of \cite{gs3}. The computation is once again carried out for both sets of boundary conditions for the condensation operator as mentioned earlier. The scalar curvature is computed next and the temperature at which the scalar curvature diverges is said to be the critical temperature in this approach. This temperature is then compared with that obtained from the matching method. The analysis gives us yet another way of comparing the results with those obtained from other analytical techniques, namely, the Sturm-Liouville eigenvalue method \cite{siop}, \cite{cai1}-\cite{dg2} and the matching method \cite{siop},\cite{hs8},\cite{fth1},\cite{sgm}.
%%%%%%%%%%%%%%%%%%%%%%%%%%%%%%%%%%%%%%%%%%%%%%%%%%%%%%%%%%%%%%%%%%%%%%%%%%%%%%%%%%%%%%%%%%

%%%%%%%%%%%%%%%%%%%%%%%%%%%%%%%%%%%%%%%%%%%%%%%%%%%%%%%%%%%%%%%%%%%%%%%%%%%%%%%%%%%%%%%%%% 

This paper is organized as follows. In section 2, we provide the basic set up for the holographic superconductors in the background of a $3+1$-dimensional electrically charged black hole in anti-de Sitter spacetime. In section 3, we compute the critical temperature using the matching method, where the matching has been carried out at several points between the boundary and the horizon. In section 4, we analytically obtain free energy expression in term of the chemical potential and the charge density. In section 5, we calculate the thermodynamic metric and the scalar curvature. We conclude finally in section 6.

%%%%%%%%%%%%%%%%%%%%%%%%%%%%%%%%%%%%%%%%%%%%%%%%%%%%%%%%%%%%%%%%%%%%%%%%%%%%%%%%%%%%%%%%%%%

%%%%%%%%%%%%%%%%%%%%%%%%%%%%%  Section 2     %%%%%%%%%%%%%%%%%%%%%%%%%%%%%%%%%%%%%%%%%%%%%%

\section{Basic set up }
The model for a holographic superconductor needs a complex scalar field coupled to a $U(1)$ gauge field in anti-de Sitter spacetime. In $3+1$-dimensions, the action for this reads
\begin{eqnarray}
S=\int d^{4}x \sqrt{-g} \left[ \frac{1}{2 \kappa^2} \left( R -2\Lambda \right) -\frac{1}{4}F^{\mu \nu} F_{\mu \nu} -(D_{\mu}\psi)^{*} D^{\mu}\psi-m^2 \psi^{*}\psi \right]
\label{tg1}
\end{eqnarray}
where $\Lambda=-\frac{3}{L^2}$ is the cosmological constant, $\kappa^2 = 8\pi G $, $G$ being the Newton's universal gravitational constant, $A_{\mu}$ and $ \psi $ represent the gauge field and scalar fields, $F_{\mu \nu}=\partial_{\mu}A_{\nu}-\partial_{\nu}A_{\mu}$; ($\mu,\nu=0,1,2,3,4$) is the field strength tensor, $D_{\mu}\psi=\partial_{\mu}\psi-iqA_{\mu}\psi$ is the covariant derivative.\\
Assuming that the plane-symmetric black hole metric can be put in the form
\begin{eqnarray}
ds^2=-f(r)dt^2+\frac{1}{f(r)}dr^2+ r^2 (dx^2 + dy^2)
\label{tg2}
\end{eqnarray}
where $f(r) = r^2 \left( 1 - \frac{r^{3}_{+}}{r^3} \right) $ and $r_{+}$ is the horizon radius and making the ansatz for the gauge field and the scalar field to be \cite{hs3}
\begin{eqnarray}
A_{\mu} = (\phi(r),0,0,0)~,~\psi=\psi(r)
\label{tg3}
\end{eqnarray}
leads to the following equations of motion for the matter fields \cite{dg2}
\begin{eqnarray}
\phi^{\prime \prime}(r) + \frac{2}{r} \phi^{\prime}(r) - \frac{2 q^2 \psi^{2}(r) \phi(r)}{f(r)} = 0
\label{e1}
\end{eqnarray}
\begin{eqnarray}
\psi^{\prime \prime}(r) + \left(\frac{2}{r} + \frac{f^{\prime}(r)}{f(r)}\right)\psi^{\prime}(r) + \left(\frac{q^2 \phi^{2}(r)}{f(r)^2}- \frac{m^{2}}{f(r)}\right)\psi(r) = 0
\label{e01}
\end{eqnarray}
where prime denotes derivative with respect to $r$. 
Since we shall carry out our analysis in the probe limit, we are allowed to set $q=1$ \cite{betti}.

The Hawking temperature of this black hole is 
\begin{eqnarray}
T_{h} = \frac{f^{\prime}(r_{+})}{4\pi} =\frac{3}{4\pi} r_{+} ~.
\label{tg4}
\end{eqnarray} 
This is interpreted as the temperature of the conformal field theory on the boundary. 
\noindent For the matter fields to be regular, one requires $\phi(r_+)=0$ and $\psi(r_{+})$ to be finite at the horizon. 

\noindent Near the boundary of the bulk, the matter fields obey \cite{hs8}
\begin{eqnarray}
\label{bound1}
\phi(r)&=&\mu-\frac{\rho}{r}\\
\psi(r)&=&\frac{\psi_{-}}{r^{\Delta_{-}}}+\frac{\psi_{+}}{r^{\Delta_{+}}}
\label{bound2}
\end{eqnarray}
where $ \Delta_{\pm} = \frac{3\pm\sqrt{9+4m^2 }}{2}$. 
The parameters $\mu$ and $\rho$ are interpreted to be dual to the chemical potential and charge density of the conformal field theory on the boundary.

\noindent At this point, we make change of coordinates $z=\frac{1}{r}$. Under this transformation, the metric (\ref{tg2}) takes the form
\begin{eqnarray}
ds^{2} &=& \frac{1}{z^2}\left(-F(z)dt^{2} +\frac{dz^{2}}{F(z)} + dx^{2} + dy^{2}\right) \nonumber \\
f(z) &=& \frac{1}{z^2}\left( 1 - \frac{z^3}{z^{3}_{h}} \right) = \frac{1}{z^2} F(z)
\label{tg5}
\end{eqnarray} 
where $F(z) = (1 - \frac{z^3}{z^{3}_{h}})$ and $z_{h}=\frac{1}{r_+}$~. \\
The Hawking temperature becomes $ T_{h}=\frac{3}{4\pi z_{h}}$ and the field eq.(s) (\ref{e1}),(\ref{e01}) become\footnote{There is an error in the factor of $\frac{1}{2}$ written down in front of the matter part of the action in \cite{fth3} as it does not lead to the equation of motion (\ref{tg6}) for the field $\phi$ with the correct numerical factor.}
\begin{eqnarray}
\label{tg6}
\phi^{\prime \prime}(z) - \frac{2 \psi^{2}(z)}{z^2 F(z)} \phi(z) &=& 0 \\
\psi^{\prime \prime}(z) + \left(\frac{F^{\prime}(z)}{F(z)} - \frac{2}{z}\right)\psi^{\prime}(z) + \left(\frac{\phi^{2}(z)}{F^{2}(z)}- \frac{m^{2}}{z^2 F(z)}\right)\psi(z) &=& 0
\label{tg7}
\end{eqnarray}
where prime now denotes derivative with respect to $z$. In the next section we shall employ the matching method in the interval $(0, z_h)$ to obtain the critical temperature below which the scalar field condensation takes place. The boundary condition $\phi(r_+)=0$ in $z$-coordinate translates to $\phi(z=z_h)=0$. The asymptotic behaviour of the fields read
\begin{eqnarray}
\phi(z) &=& \mu - \rho z \\
\psi(z) &=& \psi_{-} z^{\Delta_{-}}~ +~ \psi_{+} z^{\Delta_{+}}~~.
\label{tg8}
\end{eqnarray}

%%%%%%%%%%%%%%%%%%%%%%%%%%%%%%%%%%%%%%%%%%%%%%%%%%%%%%%%%%%%%%%%%%%%%%%%%%%%

%%%%%%%%%%%%%%%%%%%%%%%%  Section 3 %%%%%%%%%%%%%%%%%%%%%%%%%%%%%%%%%%%%%%%%

\section{Critical temperature from matching method}
\noindent To apply the matching method, we require the fields to be finite at the horizon. The Taylor series expansions of these fields near the horizon read
\begin{eqnarray}
\label{tg9}
\phi_{h}(z) &=& \phi(z) + \phi^{\prime}(z_h)(z-z_h) + \frac{\phi^{\prime \prime}(z_h)}{2}(z-z_h)^2 + ..........\\
\psi_{h}(z) &=& \psi(z) + \psi^{\prime}(z_h)(z-z_h) + \frac{\psi^{\prime \prime}(z_h)}{2}(z-z_h)^2 + .........
\label{tg10}
\end{eqnarray} 
To compute the undetermined coefficients, we use the boundary condition $\phi(z_h) = 0$ along with $f(z_h) = 0$ and eq.(s)(\ref{tg6}),(\ref{tg7}). This yields\footnote{There are errors in sign in the undetermined coefficients in \cite{fth3}. Also factors of $z_{h}$ are missing.}
\begin{eqnarray}
\label{tg11}
\phi^{\prime\prime}(z_h) &=& -\frac{2}{3 z_h} \phi^{\prime}(z_h) \psi^{2}(z_h) \\
\psi^{\prime}(z_h) &=& -\frac{m^2}{3 z_h} \psi(z_h)~~~ ; ~~~\psi^{\prime\prime}(z_h) = \frac{\psi(z_h)}{18 z^{2}_{h}}\left[ m^4 + 6 m^2 - z^4 \phi^{\prime 2}(z_h) \right]
\label{tg12}
\end{eqnarray}
\noindent In the rest of our analysis, we shall set $m^2 = -2$. This is consistent with the Breitenlohner-Freedman bound\cite{fth4},\cite{bf2}. Hence the near horizon expansions of these fields upto $\mathcal{O}(z^2)$ read 
\begin{eqnarray}
\label{tg13}
\phi_{h}(z) &=& \phi^{\prime}(z_h) \left[(z-z_h) - \frac{\psi^{2}(z_h)}{3 z_h}(z-z_h)^{2} \right] \\
\psi_{h}(z) &=& \psi(z_h) \left[1 + \frac{2}{3 z_h}(z - z_h) - \frac{(8 + z^{4}_{h} \phi^{\prime 2}(z_h) )}{36 z^{2}_{h}}(z-z_h)^{2} \right]~.
\label{tg14}
\end{eqnarray}
The matching method involves matching the near horizon expression of the fields with the asymptotic solution of these field at any arbitrary point between the horizon and the boundary, say $z=\frac{z_{h}}{2}$~. In our analysis, we shall match the solution at $z = \frac{z_h}{\lambda}$, where $ \lambda $ lies between $ [1 , \infty]$. We shall later on set specific values of $\lambda$. The matching conditions are 
\begin{eqnarray}
\label{tg15}
\phi_{h}\left(\frac{z_h}{\lambda}\right) = \phi_{b}\left(\frac{z_h}{\lambda}\right)  ~~~;~~~  \phi^{\prime}_{h}\left(\frac{z_h}{\lambda}\right) &=& \phi^{\prime}_{b}\left(\frac{z_h}{\lambda}\right) \\
\psi_{h}\left(\frac{z_h}{\lambda}\right) = \psi_{b}\left(\frac{z_h}{\lambda}\right)  ~~~;~~~  \psi^{\prime}_{h}\left(\frac{z_h}{\lambda}\right) &=& \psi^{\prime}_{b}\left(\frac{z_h}{\lambda}\right).
\label{tg16}
\end{eqnarray} 
From eq.(\ref{tg15}), we obtain the following relations
\begin{eqnarray}
\label{tg17a}
\psi^{2}(z_h) &=& \frac{3\lambda}{1 - \lambda^2} \left( \frac{\mu}{z_{h} \phi^{\prime}(z_h)} + 1 \right)  \\
\rho &=& \frac{\lambda}{z_{h}(\lambda + 1)}\left[ 2\mu + z_{h}\phi^{\prime}(z_h)\left(1 - \frac{1}{\lambda} \right) \right]~.
\label{tg17}
\end{eqnarray}
Similarly, from eq.(\ref{tg16}), we get 
\begin{eqnarray}
\label{tg18}
\psi_{-/+} &=& \left[ 1 - \frac{(\lambda-1)}{3\lambda}\frac{(3\lambda\Delta + 2)}{\Delta(\lambda - 1) + 2} \right]\left(\frac{\lambda}{z_{h}}\right)^{\Delta} \psi(z_h) \\
\phi^{\prime 2}(z_h) &=& \frac{1}{z^{4}_{h}}\left[ \frac{12\lambda}{(\lambda - 1)}\frac{\lambda\Delta -2 (1-\Delta)}{\Delta(\lambda -1) +2} -8 \right] ~~\Rightarrow ~ \phi^{\prime}(z_{h}) = -\frac{\chi(\lambda,\Delta)}{z^{2}_{h}}
\label{tg19}
\end{eqnarray} 
where 
\begin{eqnarray}
\chi(\lambda,\Delta) = \sqrt{\frac{12\lambda}{(\lambda - 1)}\frac{\lambda\Delta -2 (1-\Delta)}{\Delta(\lambda -1) +2} -8}~.
\end{eqnarray}
In eq.(\ref{tg18}), $\psi_{-}$ is for $\Delta= \Delta_{-} = 1 $ and $\psi_{+}$ is for $\Delta =\Delta_{+} = 2$. Note that we consider the negative sign before the square root of $\phi^{\prime}(z_h)$ because $\phi^{\prime} (z_h)$ is the electric field due to the charge of the black hole.\\
\noindent Substituting $\phi^{\prime}(z_h)$ from eq.(\ref{tg19}) in eq.(s)(\ref{tg17a}),(\ref{tg17}), we obtain
\begin{eqnarray}
\psi(z_h) &=& \sqrt{ \frac{3\lambda}{\lambda^2 - 1} \left( \frac{\mu z_h}{\chi(\lambda, \Delta)} - 1 \right)} \\
\rho &=& \frac{\lambda}{z_{h}(\lambda + 1)}\left[ 2\mu -\frac{\chi(\lambda,\Delta)}{z_h}\left(1 - \frac{1}{\lambda} \right) \right].
\end{eqnarray} 
Using eq.(s) (\ref{tg17a})-(\ref{tg19}) and $T = \frac{3}{4\pi z_h}$, we obtain the condensation operator and the critical temperature in terms of the chemical potential and the charge density
\begin{eqnarray}
\langle \mathcal{O}\rangle = \gamma_{(\mu)} T^{\Delta}_{c} \left( 1 - \frac{T}{T_{c}}\right)^{1/2} ~~~ ; ~~~ T_{c} &=& \xi_{(\mu)} \mu \\
\langle \mathcal{O}\rangle = \gamma_{(\rho)} T^{\Delta}_{c} \left( 1 - \frac{T}{T_{c}}\right)^{1/2} ~~~ ; ~~~ T_{c} &=& \xi_{(\rho)} \sqrt{\rho}
\end{eqnarray}
where
\begin{eqnarray}
\gamma_{(\mu)} = \sqrt{\frac{3\lambda^2}{\lambda^2 -1}}\left[ 1 - \frac{(\lambda-1)}{3\lambda}\frac{(3\lambda\Delta + 2)}{\Delta(\lambda - 1) + 2} \right]\left(\frac{4\pi\lambda}{3}\right)^{\Delta} ~~;~~ \xi_{(\mu)} &=& \frac{3}{4\pi\chi(\lambda,\Delta)} \\
\gamma_{(\rho)} = \sqrt{\frac{3\lambda}{\lambda -1}}\left[ 1 - \frac{(\lambda-1)}{3\lambda}\frac{(3\lambda\Delta + 2)}{\Delta(\lambda - 1) + 2} \right]\left(\frac{4\pi\lambda}{3}\right)^{\Delta} ~~;~~ \xi_{(\rho)} &=& \frac{3}{4\pi\sqrt{\chi(\lambda,\Delta)}}~.
\end{eqnarray}

\noindent We now consider $\psi_{-}= \langle \mathcal{O}\rangle, ~ \psi_{+}= 0$. As mentioned earlier setting $\psi_{+}=0$ implies that the condensate $\psi_{-}$ forms in the absence of the source term $\psi_{+}$. We also choose the matching point to be the middle point between the horizon and the boundary i.e. $z=\frac{z_h}{2}$ which implies $\lambda=2$ in the above expressions. This gives $\chi = \sqrt{8} $ for the value of $\Delta=\Delta_{-}=1$. Hence the condensation operator and the critical temperature reads
\begin{eqnarray}
\langle \mathcal{O}_{-}\rangle_{(\mu)} = \frac{80\pi}{27}T_{c} \left( 1 - \frac{T}{T_{c}}\right)^{1/2}~~;~~ T_{c} &=& 0.084 \mu  \\
\langle \mathcal{O}_{-}\rangle_{(\rho)} = \frac{40\pi\sqrt{6}}{27}T_{c} \left( 1 - \frac{T}{T_{c}}\right)^{1/2}~~;~~ T_{c} &=& 0.142 \sqrt{\rho}~.
\end{eqnarray}\\
For the other case, that is $\psi_{+}= \langle \mathcal{O}\rangle, ~ \psi_{-}= 0$, we once again choose matching point to be the middle point between the horizon and the boundary i.e. $z=\frac{z_h}{2}$. This gives $\chi = \sqrt{28} $. Hence the condensation operator and the critical temperature reads
\begin{eqnarray}
\langle \mathcal{O}_{+}\rangle_{(\mu)} = \frac{160\pi^2}{27}T^{2}_{c} \left( 1 - \frac{T}{T_{c}}\right)^{1/2}~~;~~ T_{c} &=& 0.0451 \mu  \\
\langle \mathcal{O}_{+}\rangle_{(\rho)} = \frac{80\pi^{2}\sqrt{6}}{27}T^{2}_{c} \left( 1 - \frac{T}{T_{c}}\right)^{1/2}~~;~~ T_{c} &=& 0.104 \sqrt{\rho}~.
\end{eqnarray}\\
%%%%%%%%%%%%%%%%%%%%%%%%%%%%%%%%%%%%%%%%%%%%%%%%%%%%%%%%%%%%%%%%%
%%%%%%%%%%%%%%%%%%%%%%%% %%%%%%%%%%%%%%%%%%%%%%%%
\section{Free energy of the holographic superconductor}  
We now proceed to compute the free energy at a finite temperature of the field theory living on the boundary of the $3+1$- bulk theory. To do this the holographic approach is used in relating the free energy $(\Omega)$ of the boundary field theory to the product of the temperature $(T)$ and the on-shell value of the Abelian-Higgs sector of the Euclidean action $(S_{E})$ \cite{fth2}.\\
To proceed further, we first write down the action for the action for the Abelian-Higgs sector
\begin{eqnarray}
S_{M} = \int d^{4}x \sqrt{-g} \left[-\frac{1}{4}F^{\mu \nu} F_{\mu \nu} -(D_{\mu}\psi)^{*} D^{\mu}\psi-m^2 \psi^{*}\psi \right]~.
\end{eqnarray}
Using the ansatz, $m^{2}=-2$ and $q=1$, we get
\begin{eqnarray}
S_{M}= \int d^{4}x \left[\frac{\phi^{\prime 2}(z)}{2} - \frac{F(z)\psi^{\prime 2}(z)}{z^{2}} + \frac{\phi^{2}(z)\psi^{2}(z)}{z^{2}F(z)} +\frac{2\psi^{2}(z)}{z^4} \right]~.
\end{eqnarray}
Applying the boundary condition ($\phi(z_{h}) = 0$) and the equations of motion (\ref{tg6}),(\ref{tg7}), we obtain the on-shell value of the action $S_{E}$ to be
\begin{eqnarray}
S_{o} = \int d^{3}x \left[-\frac{1}{2}\phi (z)\phi^{\prime}(z) \mid_{z=0} ~+~ \frac{F(z)\psi(z)\psi^{\prime}(z)}{z^{2}}\mid_{z=0} ~-~ \int^{z_{h}}_{0} dz \frac{\phi^{2}(z)\psi^{2}(z)}{z^{2}F(z)} \right] ~.
\end{eqnarray}
Substituting the asymptotic behaviour of $\phi(z)$ and $\psi(z)$ in the above action, we get
\begin{eqnarray}
S_{o}= \int d^{3}x \left[\frac{\mu\rho}{2} + 3\psi_{+}\psi_{-} +\left(\frac{\psi^{2}_{-}}{z}\right)\mid_{z=0} ~-~ \int^{z_{h}}_{0} dz \frac{\phi^{2}(z)\psi^{2}(z)}{z^{2}(1-z^{3}/z^{3}_{h})} \right]~.
\end{eqnarray}
Note that the term $\left(\frac{\psi^{2}_{-}}{z}\right)\mid_{z=0}$ diverges and therefore one needs to add a counter term at the boundary to cancel this divergence. This counter term comes from the counter action which reads
\begin{eqnarray}
S_{c} = -\int d^{3}x \left(\sqrt{-h} \psi^{2}(z) \right)\mid_{z=0}
\end{eqnarray} 
where $h$ is the determinant of the induced metric on the $AdS$ boundary. Using the asymptotic behaviour of $\psi(z)$ and evaluating this term, we obtain
\begin{eqnarray}
S_{c} = -\int d^{3}x \left[2\psi_{+}\psi_{-} + \left(\frac{\psi^{2}_{-}}{z}\right)\mid_{z=0} \right]~.
\end{eqnarray}\\
The free energy of the $2+1$-boundary field theory can be obtained by adding $S_o$ and $S_c$. This yields 
\begin{eqnarray}
\Omega = -T (S_{o} + S_{c}) = \beta T V_{2} \left[ -\frac{\mu\rho}{2} -\psi_{+}\psi_{-} + I~\right] 
\end{eqnarray}
where $\int d^{3}x = \beta V_{2}$, $V_{2}$ is the volume of the $2$-dimensional space of the boundary and the integral $I$ reads 
\begin{eqnarray}
I = \int^{z_h}_{0} dz \frac{\phi^{2}(z)\psi^{2}(z)}{z^{2}(1- z^{3}/z^{3}_{h})} = \int^{z_{h}/\lambda}_{0} dz \frac{\phi^{2}_{b}(z)\psi^{2}_{b}(z)}{z^{2}(1- z^{3}/z^{3}_{h})} + \int^{z_h}_{z_{h}/\lambda} dz \frac{\phi^{2}_{h}(z)\psi^{2}_{h}(z)}{z^{2}(1- z^{3}/z^{3}_{h})} \equiv \mathcal{I}_{1} + \mathcal{I}_{2}~. \nonumber \\
\end{eqnarray}
To evaluate the integral, we rewrite the matter field in the following form
\begin{eqnarray}
\psi(z_h) = \chi_{1}\sqrt{\frac{\mu z_h}{\chi} - 1}~~~;~~~ \psi_{-/+} = \chi_{2}\frac{\psi(z_h)}{z^{\Delta}_{h}}
\end{eqnarray}
where $\chi_{1} = \sqrt{\frac{3\lambda^2}{\lambda^2 -1}}$ and $\chi_{2} = \lambda^{\Delta} \left[1 - \frac{(\lambda-1)}{3\lambda}\frac{3\lambda\Delta + 2 }{\Delta(\lambda-1) +2 } \right] $.
Now using the substitution $z=z_{h} l$, we obtain
\begin{eqnarray}
\mathcal{I}_{1} = \int^{z_{h}/\lambda}_{0} dz \frac{\phi^{2}_{b}(z)\psi^{2}_{b}(z)}{z^{2}(1- z^{3}/z^{3}_{h})} &=& \int^{z_{h}/\lambda}_{0} dz \frac{(\mu - \rho z)^2 \psi^{2}_{-/+} z^{2\Delta}}{z^{2}(1- z^{3}/z^{3}_{h})} \nonumber \\ 
&=&\psi^{2}_{-/+} z^{2\Delta -1}_{h} \left[\mu^{2} \mathcal{A}_{1} + \rho^{2} z^{2}_{h} \mathcal{A}_2 - 2\mu\rho z_h \mathcal{A}_3 \right] \nonumber \\
&=& \chi^{2}_{2}\frac{\psi^{2}(z_h)}{z_h} \left[ B_{1}\mu^{2} + B_{2}\frac{\mu}{z_h} + B_{3}\frac{1}{z^{2}_{h}} \right] \nonumber \\
&=& \chi^{2}_{1}\chi^{2}_{2} \left[C_{1}\mu^{3} + C_{2}\frac{\mu^2}{z_h} + C_{3}\frac{\mu}{z^{2}_h} + C_{4}\frac{1}{z^{3}_h} \right]
\end{eqnarray}
where the constants are given by the following relations
\begin{eqnarray}
\mathcal{A}_{1} &=& \int^{1/\lambda}_{0} \frac{l^{2\Delta-2} dl}{(1-l^3)} ~~~;~~~\mathcal{A}_{2}= \int^{1/\lambda}_{0} \frac{l^{2\Delta} dl}{(1-l^3)}~~~;~~~\mathcal{A}_{3} = \int^{1/\lambda}_{0} \frac{l^{2\Delta-1} dl}{(1-l^3)} \nonumber \\
B_{1} &=& \mathcal{A}_{1}+ \frac{4\mathcal{A}_{2}\lambda^2}{(1+\lambda)^{2}} - \frac{4\mathcal{A}_{3}\lambda}{(1+\lambda)} \nonumber \\
B_{2} &=& \frac{2\chi \mathcal{A}_{3} (\lambda -1)}{(1+\lambda)} - \frac{4\chi \mathcal{A}_{2}(1-1/\lambda)}{(1+1/\lambda)^{2}} \nonumber \\
B_{3} &=& \frac{\mathcal{A}_{2}\chi^{2}(\lambda-1)^2}{(\lambda +1)^2} \nonumber \\
C_{1} &=& \frac{B_{1}}{\chi} ~~;~~ C_{2}=\frac{B_2}{\chi} -B_1 ~~;~~ C_{3}= \frac{B_3}{\chi} - B_2 ~~;~~ C_{4} = -B_3 ~~.
\end{eqnarray}
The evaluation of the integral $\mathcal{I}_{2}$ can be done in a similar way and yields
\begin{eqnarray}
\mathcal{I}_{2} = \int^{z_h}_{z_{h}/\lambda} dz \frac{\phi^{2}_{h}(z)\psi^{2}_{h}(z)}{z^{2}(1- z^{3}/z^{3}_{h})} = \phi^{\prime 2}(z_h)\psi^{2}(z_h) z_h \left[\mathcal{A}_{4} + \left(\frac{4}{3} - \frac{2\psi^{2}(z_h)}{3} \right)\mathcal{A}_{5} \right. \nonumber \\
\left. + \left(\frac{\psi^{4}(z_h)}{9} + \frac{4}{9} - \frac{8+\chi^2}{18} -\frac{8\psi^{2}(z_h)}{9} \right)\mathcal{A}_{6} + \left(\frac{4\psi^{4}(z_h)}{27} -\frac{2\psi^{2}(z_h)}{3}\left(\frac{4}{9}-\frac{8+\chi^2}{18}\right) -\frac{8+\chi^{2}}{27}\right)\mathcal{A}_{7} \right. \nonumber \\
\left. + \left(\frac{(8+\chi^{2})^2}{36^2} +\left(\frac{4}{9}-\frac{8+\chi^2}{18}\right)\frac{\psi^{4}(z_h)}{9} +\frac{2(8+\chi^2)\psi^{2}(z_h)}{81} \right)\mathcal{A}_{8} + \frac{(8+\chi^2)^{2}\psi^{4}(z_h)}{9(36)^2}\mathcal{A}_{10} \right. \nonumber \\
\left. + \left(\frac{-(8+\chi^2)\psi^{4}(z_h)}{243}-\frac{2(8+\chi^2)^{2}\psi^{2}(z_h)}{3(36)^2} \right)\mathcal{A}_{9}   \right] 
\end{eqnarray}
where the constants $\mathcal{A}_{n}$ ($n=4,~5,~6,~7,~8,~9,~10$) are given by the following relations
\begin{eqnarray}
\mathcal{A}_{n} = \int^{1}_{1/\lambda}\frac{(l-1)^{n-2} dl}{l^2 (1-l^3)} \label{zz}
\end{eqnarray} 
After simplification of the above expression, we get
\begin{eqnarray}
\mathcal{I}_{2} &=& \phi^{\prime 2}(z_h)\psi^{2}(z_h) z_{h}\left[B_4 + B_{5}\psi^{2}(z_h) +B_{6}\psi^{4}(z_h) \right] \nonumber \\
&=& \phi^{\prime 2}(z_h)\psi^{2}(z_h) z_{h}\left[C_5 + C_{6}\mu z_h  +C_{7}\mu^{2}z^{2}_{h} \right] \nonumber \\
&=& \chi^{2}_{1}\chi^{2}_{3} \left[D_{1}\mu^3 + D_{2}\frac{\mu^2}{z_h} + D_{3}\frac{\mu}{z^{2}_{h}} + D_{4}\frac{1}{z^{3}_{h}} \right]
\end{eqnarray} 
where the constants in the above expression are given by the following relations
\begin{eqnarray}
B_{4} &=& \mathcal{A}_{4} + \frac{4}{3}\mathcal{A}_{5} + \left(\frac{4}{9} - \frac{8+\chi^2}{18} \right)\mathcal{A}_{6} -\frac{8+\chi^2}{27}\mathcal{A}_{7} + \frac{(8+\chi^2)^{2}}{36^{2}}\mathcal{A}_{8} \nonumber \\
B_{5} &=& -\frac{2}{3}\mathcal{A}_{5} - \frac{8}{9}\mathcal{A}_{6} -\frac{2}{3} \left(\frac{4}{9} - \frac{8+\chi^2}{18} \right)\mathcal{A}_{7} +\frac{16+2\chi^2}{81}\mathcal{A}_{8} -\frac{2}{3} \frac{(8+\chi^2)^{2}}{36^{2}}\mathcal{A}_{9} \nonumber \\
B_{6} &=& \frac{1}{9}\mathcal{A}_{6} + \frac{4}{27}\mathcal{A}_{7} + \left(\frac{4}{9} - \frac{8+\chi^2}{18} \right)\frac{\mathcal{A}_{8}}{9} -\frac{8+\chi^2}{243}\mathcal{A}_{9} + \frac{(8+\chi^2)^{2}}{36^{2}}\frac{\mathcal{A}_{10}}{9} \nonumber \\
C_{5} &=& B_{4} - \chi^2_{1}B_{5} + \chi^{4}_{1}B_{6} ~~;~~ C_{6}= \frac{\chi^{2}_{1}}{\chi}B_{5} -\frac{2\chi^{4}_{1}}{\chi}B_{6} ~~;~~ C_{7}= \frac{\chi^{4}_{1}}{\chi^2}B_{6} \nonumber \\
D_{1} &=& \frac{C_{7}}{\chi} ~~;~~ D_{2}=\frac{C_6}{\chi} -C_7 ~~;~~ D_{3}= \frac{C_5}{\chi} - C_6 ~~;~~ D_{4} = -C_5~~.
\end{eqnarray}
Adding $\mathcal{I}_{1}$ and $\mathcal{I}_{2}$, we finally get
\begin{eqnarray}
I = E_{1} \mu^3 + E_{2} \frac{\mu^2}{z_h} + E_{3}\frac{\mu}{z^{2}_{h}} + E_{4} \frac{1}{z^{3}_{h}}
\end{eqnarray}
where 
\begin{eqnarray}
E_{1} &=& \chi^{2}_{1} \chi^{2}_{2} C_{1} + \chi^{2}_{1} \chi^{2}_{3} D_{1} ~~;~~ E_{2}= \chi^{2}_{1} \chi^{2}_{2} C_{2} + \chi^{2}_{1} \chi^{2}_{3} D_{2} \nonumber \\
E_{3} &=& \chi^{2}_{1} \chi^{2}_{2} C_{3} + \chi^{2}_{1} \chi^{2}_{3} D_{3} ~~;~~ E_{4}= \chi^{2}_{1} \chi^{2}_{2} C_{4} + \chi^{2}_{1} \chi^{2}_{3} D_{4}~.
\end{eqnarray}
Hence the analytical expression for the free energy in terms of the chemical potential reads
\begin{eqnarray}
\frac{\Omega}{V_2} &=& -\frac{\lambda}{\lambda+1}\frac{\mu^2}{z_h} + \frac{(\lambda -1)\chi}{2(\lambda +1)}\frac{\mu}{z^{2}_{h}} + I \nonumber \\
&\equiv & G_{1}\mu^3 + G_{2}\mu^2 T + G_{3}\mu T^2 + G_{4} T^3
\end{eqnarray}
where 
\begin{eqnarray}
G_{1} &=& E_{1} ~~~;~~~ G_{2}=\left(E_{2} - \frac{\lambda}{\lambda+1} \right)\frac{4\pi}{3} \nonumber \\
G_{3} &=& \left(E_{3} - \frac{(\lambda-1)\chi}{2(\lambda+1)} \right)\frac{16\pi^2}{9} ~~~;~~~ G_{4}=\frac{64\pi^3}{27} E_{4}~.
\end{eqnarray}
This expression for the free energy can also be written in terms of the charge density as 
\begin{eqnarray}
\frac{\Omega}{V_2} = H_{1}\frac{\rho^3}{T^3} + H_{2}\frac{\rho^2}{T} + H_{3}\rho T + H_{4} T^3
\end{eqnarray}
where 
\begin{eqnarray}
H_{1} &=& \frac{G_{1}}{8}\left(1+\frac{1}{\lambda}\right)^{3} \left(\frac{3}{4\pi}\right)^{3} \nonumber \\
H_{2} &=& \frac{G_{2}}{4}\left(1+\frac{1}{\lambda}\right)^{2} \left(\frac{3}{4\pi}\right)^{2} +\frac{3\chi G_{1}}{8}\left(1+\frac{1}{\lambda}\right)^{2} \left(1-\frac{1}{\lambda}\right) \left(\frac{3}{4\pi}\right) \nonumber\\
H_{3} &=& \frac{G_{3}}{2}\left(1+\frac{1}{\lambda}\right) \left(\frac{3}{4\pi}\right)+ \frac{\chi G_{2}}{2}\left(1-\frac{1}{\lambda^2}\right) +\frac{3\chi^2 G_{1}}{8}\left(1+\frac{1}{\lambda}\right) \left(1-\frac{1}{\lambda}\right)^{2} \left(\frac{4\pi}{3}\right) \nonumber\\
H_{4} &=& \frac{\chi G_{3}}{2}\left(1-\frac{1}{\lambda}\right) \left(\frac{4\pi}{3}\right)+ \frac{\chi^2 G_{2}}{4}\left(1-\frac{1}{\lambda}\right)^{2} \left(\frac{4\pi}{3}\right)^{2} +\frac{\chi^3 G_{1}}{8}\left(1-\frac{1}{\lambda}\right)^{3} \left(\frac{4\pi}{3}\right)^{3} + G_{4}. \nonumber \\
\end{eqnarray}
In the next section, we shall investigate the thermodynamic geometry of this model in the grand canonical ensemble.

\section{Thermodynamic geometry}
With the above results in hand, we now proceed to investigate the thermodynamic geometry of this holographic superconductor.
The thermodynamic metric is defined as \cite{gs1}, \cite{gs2}
\begin{eqnarray}
g_{ij} = -\frac{1}{T}\frac{\partial^2 \omega(T,\rho)}{\partial x^{i} \partial x^{j}} = -\frac{1}{T}\frac{\partial^2 \omega(T,\mu)}{\partial x^{i} \partial x^{j}}
\end{eqnarray}
where $\omega = \frac{\Omega}{V_{2}},~ x^1 = T $ and $x^2= \rho~$ or $\mu$. Hence the components of the metric in terms of $\mu$ read
\begin{eqnarray}
g_{TT} &=& -\left[2G_{3}\frac{\mu}{T} + 6G_{4} \right]  \nonumber \\
g_{T\mu} &=& g_{\mu T} = -\left[2G_{2}\frac{\mu}{T} + 2G_{3} \right] \nonumber \\
g_{\mu\mu} &=& -\left[6G_{1}\frac{\mu}{T} + 2G_{2} \right]
\label{tg58}
\end{eqnarray} 
and in terms of $\rho$ read
\begin{eqnarray}
g_{TT} &=& -\left[12 H_{1}\frac{\rho^3}{T^6} + 2 H_{2}\frac{\rho^2}{T^4} + 6 H_{4} \right]  \nonumber \\
g_{T\rho} &=& g_{\rho T} =\left[9 H_{1}\frac{\rho^2}{T^5} + 2 H_{2}\frac{\rho}{T^3} -H_{3}\frac{1}{T} \right] \nonumber \\
g_{\rho\rho} &=& -\left[6 H_{1}\frac{\rho}{T^4}  + 2 H_{2} \frac{1}{T^2}\right]~.
\label{tg59}
\end{eqnarray}
The scalar curvature of a general metric
\begin{eqnarray}
ds^{2}_{th} = g_{11} (dx^{1})^{2} + 2 g_{12} dx^{1} dx^{2} + g_{22} (dx^{2})^2
\label{tg60}
\end{eqnarray}
is given by \cite{gs3}
\begin{eqnarray}
R =\frac{-1}{\sqrt{g}}\left[ \frac{\partial}{\partial x^{1}}\left(\frac{g_{12}}{g_{11}\sqrt{g}}\frac{\partial g_{11}}{\partial x^{2}} - \frac{1}{\sqrt{g}}\frac{\partial g_{22}}{\partial x^{1}}\right) + \frac{\partial}{\partial x^{2}}\left(\frac{2}{\sqrt{g}}\frac{\partial g_{22}}{\partial x^{2}} - \frac{1}{\sqrt{g}}\frac{\partial g_{11}}{\partial x^{2}} -\frac{g_{12}}{g_{11}\sqrt{g}}\frac{\partial g_{11}}{\partial x^{1}}\right) \right]~.
\label{tg61}
\end{eqnarray} 
To look for any singularity in $R$, one has to see whether the denominator of the right hand side of eq.(\ref{tg61}) vanishes. The condition of the divergence of $R$ is $ det g_{ij} = 0$. For the metric (\ref{tg58}), this gives
\begin{eqnarray}
4\left[(3G_{2}G_{4} -G^{2}_{3}) + (9G_{1}G_{4} -G_{2}G_{3})\frac{\mu}{T} + (3G_{1}G_{3} -G^{2}_{2})\frac{\mu^2}{T^2}\right] = 0 
\end{eqnarray}
and for the metric (\ref{tg59}), this gives
\begin{eqnarray}
\left[-9H^{2}_{1}\frac{\rho^4}{T^{10}} +18 H_{1}H_{3}\frac{\rho^2}{T^6} + (36H_{1}H_{4} +4H_{2}H_{3})\frac{\rho}{T^4} + (12H_{2}H_{4} -H^{2}_{3})\frac{1}{T^2} \right]=0~. \nonumber \\
\end{eqnarray} 
The temperature for which the scalar curvature vanishes can be obtained by solving these equations. We obtain this critical temperature for the two different cases $\Delta_{-}\neq 0~,~\Delta_{+}= 0~$ and $\Delta_{+}\neq 0~,~\Delta_{-}= 0~$ for a set of values of $\lambda$ and compare them with the results which have been obtained from the matching method. \\
\noindent If we consider $\Delta = \Delta_{+}=2$ and $\lambda = 2$ then $T_{c} = 0.084 \mu$. This does not agree with the result in \cite{fth3}. The table \ref{E4} and  \ref{E5} give the results for $\Delta=\Delta_{-}$ and $\Delta=\Delta_{+}$ obtained from the matching method and the thermodynamic geometry for a set of values of the matching point.
\begin{table}[ht]
\caption{For $\Delta = \Delta_{-}=1$, the critical temperature $T_{c}=\xi_{(\rho)}\sqrt{\rho}$ with numerical value $\xi_{(\rho)}=0.225$
\cite{siop}.}   
\centering                          
\begin{tabular}{|c| c| c| c| c| c| }            
\hline
Value of $\lambda$ & Matching point & \multicolumn{2}{c|}{From matching method} & \multicolumn{2}{c|}{From divergence of $R$} \\
\hhline{~~----}
& & $\xi_{(\mu)}$ & $\xi_{(\rho)}$ & $\xi_{(\mu)}$ & $\xi_{(\rho)}$ \\
\hline
5 & 0.20 $z_{h}$ & 0.113 & 0.164 & 0.334 & 0.709 \\ 
\hline
3 & 0.33 $z_{h}$ & 0.102 & 0.156 & 0.475 & 0.340 \\
\hline
2 & 0.50 $z_{h}$ & 0.084 & 0.142 & 0.359 & 0.345  \\
\hline
$\frac{3}{2}$ & 0.66 $z_{h}$ & 0.065 & 0.124 & 0.355 & 0.284  \\
\hline 
$\frac{5}{4}$ & 0.80 $z_{h}$ & 0.047 & 0.106 & 0.375 & 0.233 \\
\hline                
\end{tabular}
\label{E4}  
\end{table}

\begin{table}[ht]
\caption{For $\Delta = \Delta_{+} =2$, the critical temperature $T_{c}=\xi_{(\rho)}\sqrt{\rho}$ with numerical value $\xi_{(\rho)} = 0.102$
\cite{rb}.}   
\centering                          
\begin{tabular}{|c| c| c| c| c| c| }            
\hline
Value of $\lambda$ & Matching point & \multicolumn{2}{c|}{From matching method} & \multicolumn{2}{c|}{From divergence of $R$} \\
\hhline{~~----}
& & $\xi_{(\mu)}$ & $\xi_{(\rho)}$ & $\xi_{(\mu)}$ & $\xi_{(\rho)}$ \\
\hline
5 & 0.20 $z_{h}$ & 0.076 & 0.134 & 0.126 & 0.152 \\ 
\hline
3 & 0.33 $z_{h}$ & 0.060 & 0.119 & 0.099 & 0.118 \\
\hline
2 & 0.50 $z_{h}$ & 0.045 & 0.104 & 0.084 & 0.104  \\
\hline
$\frac{3}{2}$ & 0.66 $z_{h}$ & 0.033 & 0.089 & 0.080 & 0.101  \\
\hline 
$\frac{5}{4}$ & 0.80 $z_{h}$ & 0.024 & 0.076 & 0.084 & 0.102 \\
\hline                
\end{tabular}
\label{E5}  
\end{table}

\section{Conclusions}
We now summarize our findings. We obtain the value of the critical temperature and the condensation operator of a holographic superconductor living in a $2+1$-dimensions for two different sets of boundary conditions of the condensation operator using the formalism of thermodynamic geometry. The results are compared with those obtained from the matching method. The matching point is taken to be anywhere between the horizon and the boundary and the results are obtained for a set of values of the matching point. The near horizon expressions obtained by the matching method plays a crucial role in obtaining the free energy of the holographic superconductor. This in turn is used to compute the thermodynamic geometry. It is observed that the results for the critical temperature (in terms of the charge density) from the two approaches, namely, the thermodynamic geometry approach and the matching method are comparable with the numerical value when the matching point for the near horizon and the boundary behaviour of the fields is taken to be the $z=z_h/3$ between the horizon and the boundary for the $\Delta=\Delta_{-}=1$ case.
However, for the $\Delta=\Delta_{+}=2$ case, the suitable matching point where the analytical results (from the two approaches) are comparable with the numerical result is found to be $z=z_h/2$.

\section*{Acknowledgments} DG would like to thank DST-INSPIRE, Govt. of India for financial support. DG would also like to thank Prof. Biswajit Chakraborty of S.N.Bose Centre for constant encouragement.
S.G. acknowledges the support by DST SERB under Start Up Research Grant (Young Scientist), File No.YSS/2014/000180.

%%%%%%%%%%%%%%%%%%%%%%%%%%%%%%%%%%%%%%%%%%%%%%%%%%

\end{document}